\documentclass[a4wide,12pt]{article}
\usepackage[margin=1.2in]{geometry}
\usepackage[utf8]{inputenc}
\usepackage [english] {babel}
\usepackage{amsmath,amssymb}
\usepackage[numbers]{natbib}

\def \sec{\begin{section}}
\def \esec{\end{section}}

\def \be{\begin{equation}}
\def \ee{\end{equation}}

\def \beq{\begin{equation}}
\def \eeq{\end{equation}}

\def \al {\alpha}

\def \La {\Lambda}

\def \pr {\partial}
\def \ra {\rightarrow}

\def \ep {\epsilon}

\def \l {\left(}
\def \r {\right)}

\def \ll {\langle}
\def \rr {\rangle}

\DeclareMathOperator*{\Tr}{Tr}

\begin{document}
\date{}
\begin{titlepage}


\vspace{1cm}

\begin{center}
{ \Huge  Son-Yamamoto relation and Holographic RG flows}
\end{center}
\vspace{1mm}

\begin{center}
\centerline{ O. Dubinkin$^{2,3}$, A. Gorsky$^{1,2}$, and A. Milekhin$^{1,2,3}$}
\vspace{0.5cm}
\center{$^1$Institute for Information Transmission Problems, B.Karetnyi 19, Moscow, Russia \\
$^2$Moscow Institute of Physics and Technology, Dolgoprudny 141700, Russia \\
$^3$Institute for Theoretical and Experimental Physics, B.Cheryomushkinskaya 25, Moscow 117218, Russia }
\end{center}

\vspace{1cm}

\begin{center}
{\large\bf Abstract}
\end{center}
Motivated  by the Son-Yamamoto (SY) relation
which connects the three point and two-point correlators
we consider the holographic RG flows in the bottom-up approach to holographic QCD via
the Hamilton-Jacobi (HJ) equation with respect to the radial coordinate . It is shown that  the SY relation is diagonal with respect to the RG flow in the 5d  YM-CS model while  the RG equation acquires the inhomogeneous term in the model with the additional scalar field which encodes the chiral condensate.

\end{titlepage}

\section{Introduction}

The derivation of the RG flows in the effective field theories is a subtle issue and in particular in the Chiral Perturbation Theory
only one-loop calculations are well justified. Moreover there is no simple way to incorporate the non-perturbative
effects into  RG dynamics. Holography provides the new tool to consider this problem and one could
investigate the dependence on the radial coordinate which is related with a  RG scale. It was argued in \cite{verlinde}
( see \cite{skenderis} for the review) that the RG equation can be identified with the Hamilton - Jacobi equation when the radial AdS-like holographic coordinate
is treated as the time variable. This identification is consistent with the standard holographic recipe when the
classical action in the bulk theory serves as the generating functional for the correlators in the boundary theory.
The HJ equations in the bulk are supplemented by the Hamiltonian constraints - Gauss law for the bulk gauge theory
and the ADM constraints for the bulk gravity. The recent discussion on the relation between the  holographic RG 
and the conventional Wilsonian  RG can be found in
\cite{polchinski}.

It is important generically to derive the properties of the different objects under a renormalization. The simplest objects to study are the $\beta$-function and anomalous dimensions of the local operators. Some of them are  not renormalized due to the
conservation laws behind. More general situation concerns the RG properties of the correlators of the different local operators. If some operator corresponds to the special symmetry like the trace of the energy stress-tensor
 which corresponds to the dilatation one can derive  low-energy theorems like in \cite{shifman}. These low-energy
theorems are the simplest examples when the non-perturbative effects can be accounted for in the RG dynamics. It was shown
that QCD low-energy theorems  are fulfilled
in the holographic approach as well \cite{erdmen}.

The separate
question concerns the mixing of the operators and correlators under the RG flows. This mixing can be quite complicated
and matrix of anomalous dimensions of the local operators can have a huge dimension.  Sometimes, say in  $N=4$ SYM, the diagonalization of the matrix of the one-loop anomalous dimensions turns out to be equivalent to the evaluation of the spectrum of some integrable system which follows from the special symmetries of the dilatation operator. When we consider the RG properties of the multipoint correlators the situation is more involved. Roughly speaking one can use
the OPE of the local operator first then consider the RG behavior of the emerging sum of the local operators and coefficient functions and finally try to collect them back into the form of the initial correlators. Nobody guarantee that it will acquires the form of the initial correlator since the anomalous dimensions of the local operators generically
do not know about each other.

In this Letter we consider the behavior of the simplest correlators under the non-perturbative RG motivated by the Son-Yamamoto relation
derived in the holographic setting in 5d YM-CS model  \cite{sy}. This relation connects three point and two - point correlators and crucially involves the axial anomaly. If true this relation would provide the highly nontrivial
anomaly matching conditions for the resonances. It is puzzling  how it could be obtained
purely in the field theory framework without any appealing to holography and the potential Ward identity behind it
has not been identified yet. Moreover even its status is quite controversial. It reproduces the
Vainshtein relation for the magnetic susceptibility of  the chiral condensate \cite{Vainshtein} and is
fulfilled in the Chiral Perturbation Theory for two  flavors in the leading chiral logs \cite{gkkv} however
there are explicit examples when it does not work \cite{sy,col}. Since its derivation in \cite{sy} is a bit
tricky it is important to recognize its role in  more general setting. It is interesting to understand what is the principle behind this relation when it works and what gets wrong with it in the models when it is not true.

To this aim we shall focus at the RG behavior of the SY relation in the framework of the holographic HJ equation. Due to the 5d CS term the canonical momentum of the gauge field in the Hamiltonian picture gets modified and we shall take the anomalous shift into account. We shall
consider the HJ equation supplemented by the Gauss law constraint in the simplest holographic models for QCD which can be thought 
of as a generalizations of the Chiral Lagrangian when the whole tower of mesons is taken into account. We shall demonstrate that in 
the simplest model with the gauge fields in the bulk  the SY relation is diagonal with respect to non-perturbative RG  
flow  generated by HJ equation. However when  condensate is taken into account the 
inhomogeneous term arises in the RG equation which means that the SY relation can not be true at all 
scales. We shall also found the two-point correlator  diagonal under the non-perturbatibe  RG flow .

The Letter is organized as follows. First we briefly remind the models under consideration and the HJ approach to the non-perturbative RG evolution. 
Then we will demonstrate by explicit calculation that the SY relation is diagonal under RG flow in 5d YM-CS model but 
gets mixed with another correlators in the model with additional scalar field. Some directions for the future research are summarized in the Conclusion.

\sec{Bottom-up models of holographic QCD}

In this paper following \cite{sy} we will consider two holographic QCD models.
The first model \cite{hardwall} deals with the bulk Yang-Mills-Chern-Simons  theory  where chiral
symmetry breaking is incorporated by boundary condition on the additional scalar field. The model involves  vector fields $A_{L}=A_L^a t^a$, $A_{R}=A_R^a t^a$, 
where $t^a$ are generators of the algebra $u(N_f)$, 
which are dual to left and right quark
currents $j_L=j_V-j_A$, $j_R=j_V+j_A$ and
scalar field $X$ whose boundary value is related  to chiral condensate $\ll \bar q  q \rr$. The action reads as
\be
S_{YMX}=\int d^5x \sqrt{g} \Tr \l |D_m X|^2 + 3|X|^2 - \cfrac{1}{4 g_5^2} \l F_L^2 + F_R^2 \r \r
\ee
where $D_m X=\pr_m X + i(A_{R m}-A_{L m})X$ and $F_{L,R m n}=\pr_m A_{L,R n}-\pr_n A_{L,R m}-i[A_{L,R m}, A_{L,R n}]$.
In order to reproduce the chiral anomaly we add the Chern-Simons term:
\be
S=S_{YM}(A_L,A_R)+S_{CS}(A_L)-S_{CS}(A_R)
\ee
with
\be
S_{CS}(A)=\kappa \Tr \l A F^2 - \cfrac{i}{2}A^3 F - \cfrac{1}{10}A^5 \r,\ \kappa = -\cfrac{N_c}{24 \pi^2}
\ee
The expectation value of the scalar field is fixed in the chiral limit by the solution to the classical 
equation of motion
\be
X(z)= \cfrac{\sigma z^3}{2}
\ee
where $\sigma$ is proportional to the chiral condensate.
We assume the $AdS_5$ metric in the bulk theory:
\be
ds^2=\frac{1}{z^2}(-dz^2+\eta_{\mu\nu}dx^\mu dx^\nu)
\ee

In our notations Latin letters denote five-dimensional coordinates and we raise and lower them using AdS metric,
whereas Greek letters are used for four-dimensional objects and we manipulate with them using Minkowski metric $\eta_{\mu \nu}$.
Physical 4D world is located at the "UV" boundary of $AdS$ space $z=\ep \ra 0$. Also, the theory needs "IR" boundary located at $z=z_m \approx 1/\La_{QCD}$.
Below we will consider the following IR boundary conditions \cite{hardwall}:
\beq
\label{1_ir}
\pr_z A_{A \mu} = \pr_z A_{V \mu} = 0
\eeq

In the second model there is no scalar field and the chiral symmetry breaking occurs due to different boundary conditions for $A_L$ and $A_R$.  The IR brane is located at
$z=0$ and UV brane is located at $z=z_0$. The action reads as \cite{sy}:
\begin{equation}
S=\frac{1}{2}\int d^4x \int_{z_0} dz\left\{f^2(z) \Tr (F^2_{L z\mu}+F^2_{R z\mu})-\frac{1}{2g^2(z)} \Tr (F^2_{L\mu\nu}+F^2_{R\mu\nu})\right\} + S_{CS}
\end{equation}
Following Son and Yamamoto we assume $f(z)$ and $g(z)$ to satisfy the following conditions: $f(-z)=f(z)$ and
$g(-z)=g(z)$.
It is more convenient to work with vector and axial gauge connections:
\be
A_{L\mu}=V_\mu+A_\mu,\quad A_{R\mu}=V_\mu-A_\mu
\ee
which obey  the Neumann and the Dirichlet boundary conditions respectively:
\be
\label{2_ir}
\pr_z V_\mu(z=0)=0,\ A_\mu(z=0)=0
\ee
This model suffers from the following problem \cite{gk}: three-point correlation $\ll VVA \rr$ does not vanish when one of the momenta of vector fields tends to zero.
We can add a surface term to the action to resolve this problem, which leads us to the expression for the CS term \cite{gk}(in the gauge $A_z=V_z=0$):
\beq
S_{CS}=\int d^4x dz \ \l 4\kappa\epsilon^{z \alpha\beta\gamma\lambda\eta} \Tr
(3 A_{\alpha}\cfrac{F_{V\beta\gamma}}{2}F_{Vz \lambda} +  A_{\alpha}\cfrac{F_{A\beta\gamma}}{2}F_{Az \lambda} )  \r
\eeq

\esec

\sec{Hamilton-Jacobi equation}

The standard way to evaluate correlation functions using holography is to solve equations of motion in the five-dimensional
bulk and vary the on-shell action with respect to boundary conditions.  However, from classical mechanics and field theory 
we know that there is an alternative approach, namely the Hamilton-Jacobi equation which sometimes works
more effectively.

Suppose we have a 5D holographic model. It means that we deal with a five-dimensional geometry and five-dimensional bulk action $S_{5D}$.
Physical 4D world lies at the UV boundary, whose 5th coordinate we will denote by $\ep$ which can be thought of as  a UV cut-off.
If we are interested in 4D correlators of
fields $j_\al$ (where $\al$ just enumerates fields ), then, according to the holographic picture, we have to insert corresponding sources $O_\alpha$ in the
4D action:
\beq
S_{4D}[O]=\int d^4x \ \l \mathcal{L}_{4D} + \sum_\al j_\al O_\al \r
\eeq
In the proper limit we have to solve \textit{classical} equations of motion in the bulk with the fixed values of $O_\al$ at the physical UV boundary.  Then 4D quantum
 generating function equals to
\begin{equation}
 Z_{4D}[O_{boundary}]=\exp(iS_{5D}^{on-shell}[O_{boundary}])
\end{equation}
In this approach the boundary values of $O_\alpha$ play the role of a classical background chemical potentials for $j_\al$.
Below we will drop indices 5D and "on-shell" for S. In 5D classical field-theory one can switch to the Hamiltonian description in which we will
trade $\ep$ to be the "time". We introduce canonical momenta:
\begin{equation}
 \pi_\al = \cfrac{\pr \mathcal{L}}{\pr(\pr_\ep O_\al)} = \cfrac{\delta S}{\delta O_\al}
\end{equation}
where we vary the on-shell action with respect to the boundary value of $O_\alpha$. Hamiltonian is given by the Legendre transform:
\begin{equation}
 H(\pi_\al,O_\al,\ep)=\sum_\al \pi_\al \pr_\ep O_\al - \mathcal{L}
\end{equation}
and the general form of the Hamilton-Jacobi equation reads as
\begin{equation}
 \frac{\partial S}{\partial \ep}+H(\cfrac{\delta S}{\delta O_\al},O_\al,\ep)=0
\end{equation}
The advantage of the HJ equation is that we can obtain a  hierarchy of equations for correlators if we vary the HJ equations with
respect to $O_\al$, since $\ll j_\al \rr=\cfrac{\delta S}{\delta O_\al}$.

It is instructive to obtain two-point functions for models from the previous section using this method. For simplicity, let us calculate
$\ll V V \rr$ for the first model. Neglecting the axial part, we arrive at the Hamiltonian:
\be
H=-\cfrac{1}{2} g_5^2 \ep \int d^4x \l \cfrac{\delta S}{\delta A_{V \mu}(x)}  \r + \cfrac{1}{4 g_5^2 \ep} \Tr \int d^4x F_{\mu \nu}(x)^2
\ee
hence the HJ equation reads as
\be
\cfrac{\pr S}{\pr \ep}-\cfrac{1}{2} g_5^2 \ep \int d^4x \l \cfrac{\delta S}{\delta A_{V \mu}(x)}  \r + \cfrac{1}{g_5^2 \ep} \Tr \int d^4x F_{\mu \nu}(x)^2=0
\ee
If we assume that the correlation function is not too singular at the limit $\ep \ra 0$, so that $\ep \ll j_V j_V \rr \ra 0$, then after varying twice
w.r.t to the $V_A$, near the UV boundary we have
\be
\cfrac{\pr \ll j^a_{V \mu}(-p) j^b_{V \nu} (p) \rr}{\pr \ep}=-\cfrac{1}{2 g_5^2 \ep}(p^2 \eta_{\mu \nu}-p_\mu p_\nu) \delta^{ab}
\ee
therefore
\be
 \ll j^a_{V \mu}(-p) j^b_{V \nu} (p) \rr = - \cfrac{1}{2 g_5^2} \log(p \ep) (p^2 \eta_{\mu \nu}-p_\mu p_\nu) \delta^{ab}
\ee
which exactly coincides with the result found in \cite{hardwall}.

Now let us discuss boundary conditions for the Hamilton-Jacobi equation. Since we deal with the first-order differential equation, to specify boundary conditions
we need to know the values of correlators at the particular $z$. To this end, consider the bulk action:
\beq
S=\int_{z_{ir}}^{z_{uv}} dz d^4x \ \mathcal{L}
\eeq
If we take the limit $z_{uv} \ra z_{ir}$, then naively we have
\beq
S \approx (z_{ir}-z_{uv}) \int d^4x \ \mathcal{L}(z=z_{uv})
\eeq
And taking variations with respect to the boundary UV values is exceptionally simple. So far everything was applicable for both models, so
we did not specify $z_{ir}, z_{uv}$ and $\mathcal{L}$. However, we should be careful with terms like $\pr_z A_{V \mu}$. 

In the first model we have Neumann boundary conditions (\ref{1_ir}), therefore in the leading order in $z_{ir}-z_{uv}=z_m-\ep$ we can neglect $\pr_z A_{V \mu}$
and $\pr_z A_{A \mu}$. So we are left with
\beq
S=(z_m-\ep) \Tr \int d^4x \l -\cfrac{1}{4 g_5^2 \ep} \l F_{L \mu \nu}^2 + F_{R \mu \nu}^2 \r +  \cfrac{3}{\ep^3} A_{A \mu}^2 |X|^2  \r
\eeq

In the second model we consider the limit $z_0 \ra 0$. We have different boundary conditions for $A_\mu$ and $V_\mu$ - see eq. (\ref{2_ir}).
Again we can neglect $\pr_z V_\mu $. However we can no longer neglect $\pr_z A_\mu$: since at the UV boundary the value of $A_\mu$ can
be arbitrary but at the IR boundary it must be zero, we have a very sharp transition $\pr_z A_\mu = \cfrac{1}{z_0} A_\mu+O(1)$. Now
it is straightforward to write down the leading terms in the Lagrangian:
\beq
\label{2_ba}
S=z_0 \int d^4x \Tr \l \cfrac{f^2}{z_0^2} A_\mu^2 - \cfrac{1}{2g^2}  F_{V \mu \nu}^2   \r
\eeq
where the additional non-abelian terms from the CS term are omitted.

\esec

\sec{Son-Yamamoto relation}

The Son-Yamamoto relation \cite{sy} connects three-point function and two-point functions in the models described above. Let us introduce usual notations:
\begin{eqnarray}
<V^{a \bot}_\mu(q) V^{b \bot}_\nu(-q)> = \delta^{ab} \Pi^\bot_{\mu \nu}(q) q^2 \Pi_V(q) \\ \nonumber
<A^{a \bot}_\mu(q) A^{b \bot}_\nu(-q)> = \delta^{ab}  \Pi^\bot_{\mu \nu}(q) q^2 \Pi_A(q) \\ \nonumber
<V_\mu^a(k) V_\nu^{\bot b}(-k-Q) A_\alpha^{\bot c}(Q)> = \cfrac{Q^2}{4 \pi^2} \epsilon_{\mu \nu \alpha \beta} k^\beta w_T(Q) \Tr(t^a t^b t^c) , \ k \ra 0 \\ \nonumber
\Pi^\bot_{\mu \nu}(q) = \eta_{\mu \nu} - \cfrac{q_\mu q_\nu}{q^2} \\ \nonumber
\Pi^\|_{\mu \nu}(q)=\cfrac{q_\mu q_\nu}{q^2}
\end{eqnarray}
Slightly abusing notations we use the same letter for quark currents and their holographic duals.
We start from the 5d Yang-Mills action for the second model:
\begin{equation}
S=\int d^4xdz \Tr \l f^2(z)((\partial_zA_\mu)^2+(\partial_zV_\mu)^2)+\frac{1}{2g^2(z)}(F^2_{A\mu\nu}+F^2_{V\mu\nu})+12\kappa\epsilon^{z \alpha\beta\gamma\lambda\eta}
A_{\alpha}\cfrac{F_{V\beta\gamma}}{2}F_{Vz \lambda} \r
\end{equation}
and omit the term $A F_A F_A$ that makes no contribution to the 3-point correlator $\langle VVA\rangle$.
Introducing the ansatz for the bulk fields (index $0$ indicates boundary value):
\begin{eqnarray}
V_\mu(z,q) = V_\mu^{0 \bot}(q) V(z) + V_\mu^{0 \|}(q) \psi_V(z) \\ \nonumber
A_\mu(z,q) = A_\mu^{0 \bot}(q) A(z) + A_\mu^{0 \|}(q) \psi_A(z)
\end{eqnarray}
we recover the results found in \cite{sy}:
\begin{eqnarray}
\Pi_V=\cfrac{2}{q^2}f^2 V'(z_0) \\ \nonumber
w_T = \cfrac{48 \kappa}{q^2} \int_0^{z_0} A V' dz
\end{eqnarray}

Now our aim is to reproduce Son-Yamamoto's relation for transversal part of a triangle anomalies taking variational
derivatives of Hamilton-Jacobi equation.
The Hamiltonian is given by:
\be
H=\pr_z A_\mu \pi_{A}^{\mu}+\pr_z V_\mu \pi_{V}^{\mu}-\mathcal{L}
\ee
where $\pi_{A\mu}=\cfrac{\partial L}{\partial(\partial_z A_\mu(z,x))}$ and respectively for $\pi_{V\mu}$.
 Let us rewrite the last expression for $H$ in terms of $A_\mu$, $V_\mu$, $\pi_{A\mu}$ and $\pi_{V\mu}$. First of all we need to obtain an exact expression for canonical momenta:
\be
\pi_{A}^\mu=\frac{\partial L}{\partial(\partial_z A_\mu(z,x))}=2f^2(z)\partial^zA^\mu
\ee
\be
\pi_{V}^\mu=\frac{\partial L}{\partial(\partial_z V_\mu(z,x))}=2f^2(z)\partial^zV^\mu+6\kappa\epsilon^{\alpha\beta\gamma z\mu}A_\alpha F_{V\beta\gamma}
\ee
which yields the expressions for $\pr_z V_\mu$ and $\pr_z A_\mu$:
\be
\partial_z A_\mu(z,x)=\frac{\pi_{A\mu}}{2f^2(z)}
\ee
\be
\partial_z V^a_\mu(z,x)=\frac{\pi^a_{V\mu}}{2f^2(z)}-3\frac{\kappa}{f^2(z)}\epsilon^{z \alpha\beta\gamma \mu}A^b_\alpha F^c_{V\beta\gamma} \Tr(t^a t^b t^c)
\ee
Now we are able to  write down the resulting Hamilton-Jacobi equation:
\begin{equation}
\begin{split}
\begin{gathered}
\frac{\partial S}{\partial z_0}+\Tr \int d^4x\bigg(\frac{1}{2f^2}\pi_{A}^\mu\pi_{A\mu}+\frac{1}{2f^2}\pi_{V}^\mu(\pi_{V\mu}-
6\kappa\epsilon^{\alpha\beta\gamma}_{z\mu}A_\alpha F_{V\beta\gamma})-\\-\bigg(f^2(z)((\partial_zA_\mu)^2+(\partial_zV_\mu)^2)+
\frac{1}{2g^2(z)}(F^2_{A\mu\nu}+F^2_{V\mu\nu})+6\kappa\epsilon^{z \alpha\beta\gamma\lambda}A_{\alpha}F_{V\beta\gamma}F_{Vz \lambda}\bigg)\bigg)
\end{gathered}
\end{split}
\end{equation}
which can be presented in the following form 
\begin{eqnarray}
\frac{\partial S}{\partial z_0}+\int d^4x\left\{\frac{1}{4f^2}\pi_{A\mu}^2+\frac{1}{4f^2}(\pi_{V\mu}-\phi_{V\mu})^2-
\frac{1}{2g^2}(F^2_{A\mu\nu}+F^2_{V\mu\nu})\right\} \\ \nonumber 
\phi_{V\mu}=6\kappa\epsilon^{z \alpha\beta\gamma \mu}A_\alpha F_{V\beta\gamma} \qquad 
{\tilde \phi}_{V\mu}=6\kappa\epsilon^{z \alpha\beta\gamma \mu}A_\alpha {\tilde F}_{V\beta\gamma}
\end{eqnarray}
We omitted $u(N_f)$ indices for brevity, the notation should be self-evident. 

Turn now to the discussion on the RG properties of the multipoint correlators. First
let us derive the two-point correlator diagonal with respect to the RG flow.
Taking the second variational derivative of the HJ equation after the simple
algebra we get 
\be
\label{hj_2p}
\cfrac{\pr}{\pr z_0}(\Pi_A-\Pi_V) = -\cfrac{q^2}{2 f^2} (\Pi_A^2 - \Pi_V^2)
\ee
We see that the difference between the axial and vector correlators is
diagonal and the sum of these correlators defines the  $q^2$ dependent 
"anomalous dimension".

Consider now the three-point functions involving one axial and two vector currents.
Taking the Fourier transform of HJ  equation and applying the following
variation: $\cfrac{\delta^3}{\delta V_\eta(k)\delta V_\chi(-q-k)\delta A_\lambda(q)}$, we obtain
\begin{equation}
\begin{split}
\begin{gathered}
\frac{\partial }{\partial z_0}\frac{\delta^3S}{\delta V_\eta(k)\delta V_\chi(-q-k)\delta A_\lambda(q)}+\frac{1}{2f^2}\int d^4p\bigg[\langle A_\lambda A_\mu\rangle(q,p)\langle V_\eta V_\chi A_\mu\rangle(k,-q-k,-p)+\\+\langle V_\eta A_\lambda V_\mu\rangle(k,q,p)\langle V_\chi V_\mu\rangle(-q-k,-p)-\frac{\delta^2\tilde{\phi}_{V\mu}}{\delta V_\eta\delta A_\lambda}(k,q,p)\langle V_\chi V_\mu\rangle(-q-k,-p)\bigg]
\end{gathered}
\end{split}
\end{equation}
Note that we set boundary values $A_0$ and $V_0$ to zero. 

Let us write down every term in some  details:
\be
\langle A_\lambda A_\mu\rangle(q,p)=\delta(p+q)q^2 \Pi^\bot_{\lambda\mu}\Pi_A\ ,\quad \langle V_\chi V_\mu\rangle(-q-k,-p)=\delta(k+q+p)p^2 \Pi^\bot_{\chi\mu}\Pi_V
\ee
\be
\langle V_\eta V_\chi A_\mu\rangle(k,-q-k,-p)=\frac{p^2}{4\pi^2} \Pi^{\alpha\bot}_\mu(\Pi^{\beta\bot}_\chi w_T+\Pi^{\beta\parallel}_\chi w_L)\epsilon_{\alpha\beta\gamma\eta}k^\gamma\delta(p+q)
\ee
\be
\langle V_\eta A_\lambda V_\mu\rangle(k,q,p)=\frac{p^2}{4\pi^2}\Pi^{\alpha\bot}_\lambda(\Pi^{\beta\bot}_\mu w_T+\Pi^{\beta\parallel}_\mu w_L)\epsilon_{\alpha\beta\gamma\eta}k^\gamma\delta(p+q+k)
\ee
and for the Fourier transform of  $\tilde{\phi}_{V}$ we get :
\be
\tilde{\phi}_{V}(p)=6\kappa\epsilon^{z \alpha\beta\gamma \mu}\int dq'A_\alpha(q')\tilde{F}_{V\beta\gamma}(-q'-p)
\ee
\be
\frac{\delta^2\tilde{\phi}_{V\mu}}{\delta V_\eta\delta A_\lambda}(k,q,p)=12\kappa\epsilon^{z \alpha\beta\gamma \mu} k_\beta
\ee
Now we contract Hamilton-Jacobi equation with $q^\eta$ to get rid of longitudinal part of 3-point correlator and arrive at equation for $w_T$ :
\begin{equation}
q^\eta\frac{\partial }{\partial z_0}\frac{\delta^3S}{\delta V_\eta\delta V_\chi\delta A_\lambda}+
\frac{1}{2f^2}\left[\frac{q^4}{4\pi^4}(\Pi_A+\Pi_V)w_T\epsilon_{\lambda\chi\beta\eta}k^\beta q^\eta+12\kappa q^2\Pi_V\epsilon^{\lambda\chi\beta\eta}k_\beta q^\eta\right]=0
\end{equation}
This expression can be presented in the following form suitable for the discussion of SY relation
\begin{equation}
\frac{q^2}{4\pi^2}\frac{\partial w_T}{\partial z_0}+\frac{1}{2f^2}\left[\frac{q^4}{4\pi^2}(\Pi_A+\Pi_V)w_T+6\kappa q^2(\Pi_A+\Pi_V)+6\kappa q^2(\Pi_V-\Pi_A)\right]=0
\end{equation}

Recall that the original Son-Yamamoto relation reads as \cite{sy}:
\be
\label{sy}
(S-Y)=w_T-\frac{N_c}{Q^2}+\frac{N_c}{f^2_\pi}(\Pi_A-\Pi_V) = 0
\ee
where
\beq 
\label{def_f}
\cfrac{1}{f_\pi(z_0)^2}=\cfrac{1}{2} { \displaystyle \int_0^{z_0} } dz \cfrac{1}{f(z)^2}
\eeq
If we take into account the diagonal two-point correlator and substitute it 
into the variation of the HJ equation we  obtain:
\be
\label{victory}
\cfrac{\pr}{\pr z_0} (S-Y) = -\cfrac{(\Pi_A+\Pi_V)q^2}{2f^2}(S-Y)
\ee
therefore in this model SY relation is diagonal under the renormalization group. In order to prove that the $(S-Y)$ is zero for all $z_0$, let us
take the limit $z_0 \ra 0$. Recalling (\ref{2_ba}) we see that $w_T \ra 0$ and
\be
\cfrac{N_c}{f_\pi^2}(\Pi_A-\Pi_V) \ra \cfrac{N_c}{q^2}
\ee
Hence the SY relation indeed holds at this limit and  therefore, it holds for all $z_0$.

In the first model with the additional scalar field we have  similar HJ equations:
\be
\label{1_2p}
\cfrac{\pr }{\pr \ep} (\Pi_A-\Pi_V) = \cfrac{g_5^2 \ep q^2}{2}(\Pi_A^2-\Pi_V^2) + \cfrac{3}{\ep^3}|X|^2
\ee
\be
\frac{q^2}{4\pi^2}\frac{\partial w_T}{\partial z_0}-\cfrac{g_5^2 \ep}{2}\left[\frac{q^4}{4\pi^2}(\Pi_A+\Pi_V)w_T+6\kappa q^2(\Pi_A+\Pi_V)+6\kappa q^2(\Pi_V-\Pi_A)\right]=0
\ee
And
\be
\cfrac{\pr (S-Y)}{\pr \ep} = (\Pi_A+\Pi_V)\cfrac{q^2 g_5^2 \ep}{2} (S-Y) + \cfrac{3 N_c}{\ep^3 f_\pi^2} |X|^2
\ee
Note that in this equation $f_\pi$ is defined as in the second model, see eq. (\ref{def_f}).

Therefore we see that in the model with scalar the RG equation for SY relation acquires the inhomogeneous term
which corresponds to its failure. 
Potentially in order to make this expression diagonal and look for the modified diagonal correlator one  could add a term $\theta$ to the SY relation which has to satisfy the following equation:
\be
\cfrac{\pr \theta}{\pr \ep} = \theta(\Pi_A+\Pi_V) \cfrac{q^2 g_5^2 \ep}{2}- \cfrac{3N_c}{\ep^3 f_\pi^2} |X|^2
\ee
However we have not found its proper operator realization.

It is interesting to note that  eq. $(\ref{hj_2p})$ tells us that the flow for $\Pi_A-\Pi_V$ is diagonal, but we do not expect $\Pi_A=\Pi_V$. In the second
model the problem is due to the boundary conditions: $\Pi_A=\Pi_V$ does not hold for $z_0=0$.
In the  first model the chiral symmetry is broken only by the quark condensate $X$( see eq. (\ref{1_2p})).
If it vanishes then we indeed have $\Pi_A=\Pi_V$.

\section{Conclusion}

In this Letter we have examined the behavior of SY relation under the non-perturbative RG flow. We have found that when SY relation is fulfilled
it is diagonal under the action of RG flow generated by HJ equation while when it is not true the inhomogeneous term
in the RG equation is present. We expect the the diagonal evolution under the RG flows should be one of the guiding principles in searches of more complicated anomaly matching conditions.  It would be interesting  to get the  higher correlators diagonal under RG flow in the holographic models of QCD. However we have seen that the issue of the diagonalization
is model dependent in the holographic models of QCD hence it is important to perform the similar consideration for the models when the bulk dual theory
is uniquely defined like in SUSY gauge theories. It would be interesting to formulate the diagonalization of the correlators in terms
of the string worldsheet theory. The diagonalization of the matrix of the anomalous dimensions corresponds to the
diagonalization of the spin chain Hamiltonian which arises upon the discretization of the worldsheet sigma-model. It would be interesting to formulate the problem of the diagonalization of correlators in the similar manner.

It would be also interesting to investigate a few  related problems. First of all the similar diagonalization problem can
be discussed in the gravity sector of the bulk theory when the Wheeler-de-Witt equation plays the role of the HJ equation. The second 
question concerns the baryonic sector of the theory. The baryons correspond to the instantons in the bulk theory \cite{ss} hence an 
interesting question concerns the solution to the HJ equations in the sector with non-vanishing topological charge. 
Finally it would be interesting to find the similar diagonal correlators from the HJ equation in the holographic bulk  descriptions
of the condensed matter models.

\section*{Acknowledgment}
We would like to thank A. Krikun and A. Vainshtein for the useful discussions. The work was supported in part by grants RFBR-12-02-00284
and PICS-12-02-91052. The work of A.M. was also supported by the Dynasty fellowship program. A.G. and A.M. gratefully acknowledge
support from the Simons Center for Geometry and Physics, Stony Brook University where part of the research was performed.

\esec

\bibliographystyle{unsrt}

\end{document}